\documentclass[prd,reprint,superscriptaddress,floatfix,nofootinbib]{revtex4-2}

\usepackage{amsmath,amssymb,amsfonts,amsthm}
\usepackage{graphicx}
\usepackage{enumitem}
\usepackage{bm}
\usepackage{rotating}
\usepackage{hyperref}
\usepackage{pbox}
\usepackage{array}
\usepackage{mathtools}
\usepackage{leftidx}
\usepackage{lmodern}
\usepackage[normalem]{ulem}
\usepackage{braket}
\usepackage{tensor}
\usepackage[usenames,dvipsnames]{color}
\usepackage{float}
\usepackage{footnote}
\usepackage{setspace}
\usepackage{CJKutf8}
\usepackage{dsfont}
\usepackage{mathrsfs}

\newcommand{\R}{\mathbb{R}}
\newcommand{\A}{\mathcal{A}}
\newcommand{\C}{\mathbb{C}}
\newcommand{\M}{\mathcal{M}}
\newcommand{\W}{\mathcal{W}}
\newcommand{\CS}{{C^\infty_0(\M)}}
\newcommand{\dd}{\mathrm{d}}
\newcommand{\rr}[1]{\left(#1\right)}
\newcommand{\bx}{{\bm{x}}}
\newcommand{\bk}{{\bm{k}}}
\newcommand{\sx}{\mathsf{x}}
\newcommand{\sy}{\mathsf{y}}

\newcommand{\ii}{\text{i}}
\DeclareMathOperator{\supp}{\text{supp}}
\DeclareMathOperator{\tr}{\text{Tr}}

\newcommand{\Sol}{\mathsf{Sol}}

\renewcommand{\Re}{\mathrm{Re}}
\renewcommand{\Im}{\mathrm{Im}}

\begin{document}
 
\title{Wigner function for quantum field theory via spacetime tiling}

\author{Erickson Tjoa}
\email{erickson.tjoa@uwaterloo.ca}
\affiliation{Department of Physics and Astronomy, University of Waterloo, Waterloo, Ontario, N2L 3G1, Canada}
\affiliation{Institute for Quantum Computing, University of Waterloo, Waterloo, Ontario, N2L 3G1, Canada}


\date{\today}

\begin{abstract}

We present a construction of the Wigner function for a bosonic quantum field theory that has well-defined ultraviolet (UV) and infrared (IR) properties. Our construction uses the local mode formalism in algebraic quantum field theory that is valid in any globally hyperbolic curved spacetimes, i.e., without invoking the path integral formalism. The idea is to build $N$ quantum harmonic oscillators degrees of freedom from $2N$ smeared field operators and use them to ``tile'' a Cauchy surface of the spacetime manifold. The smallest support of the smearing functions that define each local mode define the UV scale and the number of modes local modes fix the IR scale. This construction can be viewed as a form of ``covariant discretization'' of the quantum field in curved spacetimes, since the tiling of the Cauchy surface does not depend on any choice of coordinate systems or foliation. 

\end{abstract}

\maketitle


One of the main goals of relativistic quantum information (RQI) is to understand the properties of relativistic quantum field theory (QFT) using tools from quantum information theory (QIT). The interplay between QIT and QFT takes many forms, such as coupling localized particle detectors to a quantum field \cite{Tales2020GRQO,perche2022geometry,bruno2023probes,tjoa2022channel,tjoa2022fermi,Simidzija2020transmit,Landulfo2016communication,pozas2016entanglement,Gallock2021nonperturbative,tjoa2022teleport,jonsson2018transmitting,Maria2021causality}, holographic studies via Anti-de Sitter/Conformal Field Theory (AdS/CFT) correspondence \cite{blanco2013relative,hung2011holographic,casini2011towards,ryu-takayanagi2006,belin2022anything,penington2020entanglement,engelhardt2015quantum,dong2019flat,cotler2019entanglement,may2019quantum,hayden2013holographic,harper2023timelike}, and toy models of quantum gravity (e.g., using tensor networks) \cite{pastawski2015holographic,jahn2019holography,Jahn2021review,may2017tensor,hayden2007black,yoshida2017efficient,nakata2023black,bao2019beyond,hotta2018qubittoy,cao2023approximate,bao2017desitter,osborne2022desitter}. 

One of the tools in QIT that is still missing is the analog of quasiprobability distributions such as the Wigner function. In QIT, the Wigner function provides a phase space formulation of quantum mechanics \cite{wigner1932,moyal1949,Royer1977wigner,glauber1969kernel} and it takes on many roles, e.g., as a measure of non-classicality \cite{Kenfack2004nonclassical,Hyunseok2020}, non-contextuality \cite{Booth2022contextuality,Delfosse2017,gross2006wigner} and (hence) a resource for quantum computation (via its \textit{negativity}) \cite{howard2014contextuality,browne2017resource,ferrini2020simulable}. In principle, applications of QIT to relativistic settings should allow for generalizations of the Wigner functions to QFT, with possibly some suitable modifications.

Focusing on the scalar field theory, earlier attempts tried to construct the Wigner \textit{functional} \cite{calzetta1988functional,Muller1994wignerfunc,berra2020coherent,stanislaw2013fermionwigner,roux2019wigner,Zachos1999quantization}, essentially by regarding the field operator $\hat
\phi$ and its momentum $\hat{\pi}$ as the analog of position and momentum operators of a harmonic oscillator (or equivalently in terms of the momentum-dependent ladder operators $\hat{a}_\bk^{\phantom{\dagger}},\hat{a}_\bk^\dagger$), giving rise to an infinite-dimensional `optical' phase space. Since path integrals are required, the formalism has very limited computational control beyond well-known cases (whose answers are essentially known through other means or by analogy). It is also tricky to make rigorous and make sense of the (quasi-)probabilistic interpretation. The other alternative approach would be to perform a brute-force discretization of the field theory: in practice, this reduces the problem to standard continuous-variable (CV) settings with finitely-many (coupled) harmonic oscillators at the expense of losing all the relativistic features of the original QFT. The question remains as to whether there is a middle ground that retains enough relativistic features and at the same time still allows for sufficient mathematical tractability to be useful.

Here we present a construction of the Wigner function for a bosonic scalar field theory that comes with a natural ultraviolet (UV) and infrared (IR) scales. Our construction is valid in any $(n+1)$-dimensional globally hyperbolic curved spacetime $\M$ with metric $g_{ab}$. Crucially, we invoke neither the path integral formalism nor direct lattice discretization of the field theory. The idea is to build $N$ quantum harmonic oscillators degrees of freedom --- called the \textit{local modes} \cite{Ruep2021harvesting,Trevison2019partner} --- from $2N$ smeared field operators and use them to ``tile'' a Cauchy surface $\Sigma$ of the spacetime manifold $\M$. The UV scale $l_{\textsc{uv}}$ is set by the support of the spacetime smearing function that has the smallest spatial volume on the Cauchy surface $\Sigma$, i.e., $l_{\textsc{uv}} = V_{\text{min}}^{1/n}$. The IR scale $l_{\textsc{ir}}$ is set by the number of local modes $N$, since the volume of the Cauchy surface that can be ``tiled'' by the supports of the local modes is roughly $N l_{\textsc{uv}}^n$, i.e., $l_{\textsc{ir}} = N^{1/n} l_{\textsc{uv}}$. We also allow for another UV scale $\epsilon$ that arises by having each mode tile the Cauchy slice $\Sigma$ with a finite ``corridor'', a technicality that may be important in some contexts (e.g., when computing entropic quantities, see, e.g., \cite{hollands2017entanglement,Nishioka2009}). 

This construction has two important properties. First, it can be viewed as a form of ``covariant discretization'' of the quantum field, since the tiling of the Cauchy surface does not depend on any choice of coordinate systems or spacetime foliation. Furthermore, all length scales (e.g., $l_\textsc{uv},l_{\textsc{ir}}, \epsilon$) can be computed covariantly using the metric tensor. Second, the difficulty of computing the Wigner function for fixed $N$ reduces to the computation of finitely many \textit{smeared} correlation functions of the quantum field, which for the free theory can be done straightforwardly. In particular, this also implies that the Wigner function can inherit the symmetries of the QFT states which are not preserved by lattice discretization. For example, the Wigner function of the ground state of the free scalar theory is Gaussian \textit{and} Poincar\'e-invariant, the latter because the Wigner function depends only on the Wightman two-point function that is itself Poincar\'e-invariant.

\section{Wigner functions in algebraic QFT}

We will construct the Wigner function for the scalar field using the tools from algebraic QFT (see Appendix~\ref{appendix: AQFT} for details of the quantization). The idea is to construct \textit{local modes} \cite{Ruep2021harvesting,Trevison2019partner} --- essentially harmonic oscillator degrees of freedom using the smeared field operators, with spacetime smearing functions chosen such that the canonical commutation relations of the scalar field theory reduce to the Heisenberg-Weyl commutation relations.  We will retain the $\hbar$ to make contact with standard non-relativistic literature.

Given an algebra of observables $\A(\M)$ of the scalar field theory generated by \textit{smeared field operators} $\hat{\phi}(f)$, the canonical commutation relation (CCR) of the theory is given by (see Appendix~\ref{appendix: AQFT}) 
\begin{align}
    [\hat{\phi}(f),\hat{\phi}(g)] = i\hbar E(f,g)\openone\,,
    \label{eq: CCR}
\end{align}
where $E(f,g)$ is the smeared causal propagator and $f,g\in \CS$ are compactly supported smooth functions in $\M$. In particular, $E(f,g)=0$ when $f,g$ have causally disconnected supports. 

A \textit{local mode} can be constructed from a pair $f_1,f_2\in \CS$ such that $E(f_1,f_2)=1$ \cite{Ruep2021harvesting}. Effectively, we have a single mode of quantum harmonic oscillator (QHO) with basic observables
\begin{align}
    \hat{x}\coloneqq \hat{\phi}(f_1)\,,\qquad \hat{p}\coloneqq\hat{\phi}(f_2)\,,
    \label{eq: local-mode}
\end{align}
such that the CCR \eqref{eq: CCR} for the local mode  becomes the Heisenberg-Weyl algebra
\begin{align}
    [\hat{x},\hat{p}]\equiv [\hat{\phi}(f_1),\hat{\phi}(f_2)] =\ii\hbar\openone\,.
    \label{eq: oscillator-CCR}
\end{align}
The unital $*$-algebra $\mathfrak{A}$ generated by $\{\hat{x},\hat{p},\openone\}$ satisfying the Heisenberg-Weyl CCR defines a $*$-subalgebra of the full algebra of the field observables $\A(\M)$. For our purposes, it is convenient to choose $f_1,f_2$ such that $\supp(f_1)=\supp(f_2)$. This construction is not unique since there are many choices of $f_a\in \CS$ ($a=1,2$) satisfying the Heisenberg-Weyl CCR \eqref{eq: oscillator-CCR}.

From the local mode, we can define local creation and annihilation operators \cite{Ruep2021harvesting,glauber1969kernel}
\begin{equation}
    \begin{aligned}
    \hat{a} &\coloneqq \frac{1}{\sqrt{2\hbar}}(\lambda\hat{x}+\ii\lambda^{-1}\hat{p})\equiv \hat{\phi}(F_\lambda)\,,\\
    \hat{a}^\dagger &\coloneqq \frac{1}{\sqrt{2\hbar}}(\lambda\hat{x}-\ii\lambda^{-1}\hat{p}) \equiv \hat{\phi}(F^*_\lambda)\,,
    \label{eq: ladder}
    \end{aligned}
\end{equation}
where we define the complex smearing function $F_\lambda$
\begin{align}
    F_\lambda = \frac{1}{\sqrt{2\hbar}}(\lambda f+\ii \lambda^{-1}g)\,,\qquad \lambda\in \R\setminus\{0\}\,.
\end{align}
In natural units where the speed of light is $c=1$, the positive parameter\footnote{The case for $\lambda<0$ just gives us ``reflected'' operators by replacing $\hat{x}\to-\hat{x},\hat{p}\to-\hat{p}$. This does not buy us anything new due to the parity interpretation for the Wigner function by Royer \cite{Royer1977wigner}.} $\lambda>0$ has units of inverse length and can be identified with the mass and frequency of the QHO as $\lambda = \sqrt{m\omega}$. 
From the perspective of local modes, the freedom of choosing $\lambda$ amounts to rescaling $f_1\to \lambda f\,,f_2\to \lambda^{-1}f_2$ and this does not alter Eq.~\eqref{eq: oscillator-CCR}, so without loss of generality we may choose $\lambda=1$.

Next, we would like to define many local modes, equivalent to constructing $N$ QHO degrees of freedom. For this, we need $N$ pairs of spacetime smearing functions $\{(f_1^{(k)},f^{(k)}_2): k=1,2,...,N\}$ such that $f_a^{(k)} \in \CS$. Following Eq.~\eqref{eq: local-mode}, each pair defines a local mode $(\hat{x}_k,\hat{p}_k)$ given by
\begin{align}
    \hat{x}_k\coloneqq\hat{\phi}(f^{(k)}_1)\,,\quad \hat{p}_k\coloneqq\hat{\phi}(f^{(k)}_2)\,,
\end{align}
satisfying the joint Heisenberg-Weyl CCR
\begin{align}
    [\hat{x}_j,\hat{p}_k] &= \ii\hbar  \delta_{jk}\openone\,,\quad 
    [\hat{x}_j,\hat{p}_k] = [\hat{p}_j,\hat{p}_k] = 0\quad \forall j,k\,.
    \label{eq: joint-CCR}
\end{align}
As before, we assume that $\supp(f^{(k)}_1) = \supp(f^{(k)}_2)$ for all $k$. This lets us speak of the $k$-th local mode as being localized in the spacetime region $\mathcal{D}_k \coloneqq \supp(f^{(k)}_a)\subset \M$. From these we can generalize Eq.~\eqref{eq: ladder} for each mode, i.e.,
\begin{align}
    \hat{a}_k^{\phantom{k}} &\coloneqq \frac{\hat{x}_k+\ii\hat{p}_k}{\sqrt{2\hbar}}\equiv \hat{\phi}(F_k)\,,\quad F_k = \frac{f^{(k)}_1+\ii f^{(k)}_2}{\sqrt{2\hbar}}\,,
\end{align}
and similarly for $\hat{a}_k^\dagger$ and $F_k^*$. 
 
Note that the CCR in Eq.~\eqref{eq: joint-CCR} is not automatic: in order to ensure that the local modes pairwise-commute for $j\neq k$ requires some care. To achieve this, we will use the fact that in $\M$ there exists a spacelike Cauchy surface $\Sigma$ (by global hyperbolicity of $\M$). Let $O_\Sigma\subset \M$ be an open subset of $\M$ containing $\Sigma$ --- we can think of $O_\Sigma$ as the thickening of the Cauchy slice along the timelike direction. For example, in Minkowski spacetime with global coordinates $x^\mu\equiv (t,\bx)$, we can take $\Sigma$ to be the constant-$t=t_0$ surface, and for some $\epsilon>0$ we have 
\begin{align}
    O_\Sigma = \{\sx\in \M: |t(\sx)-t_0| < \epsilon\}\subset\M\,.
\end{align}
We then demand that the collection $\{f_a^{(k)}\}$ is chosen such that the supports $\mathcal{D}_k\subset O_\Sigma$ and they are pairwise spacelike. Mathematically, we are asking for 
\begin{align}
    \mathcal{D}_j\, \cap \, \supp\bigr(Ef^{(k)}_a\bigr) = \emptyset\,,
\end{align}
for all $j\neq k$, i.e., that each $(\mathcal{D}_j$ is in the \textit{causal complement} of every other $\mathcal{D}_k$.

\begin{figure}[tp]
    \centering
    \includegraphics[scale=0.4]{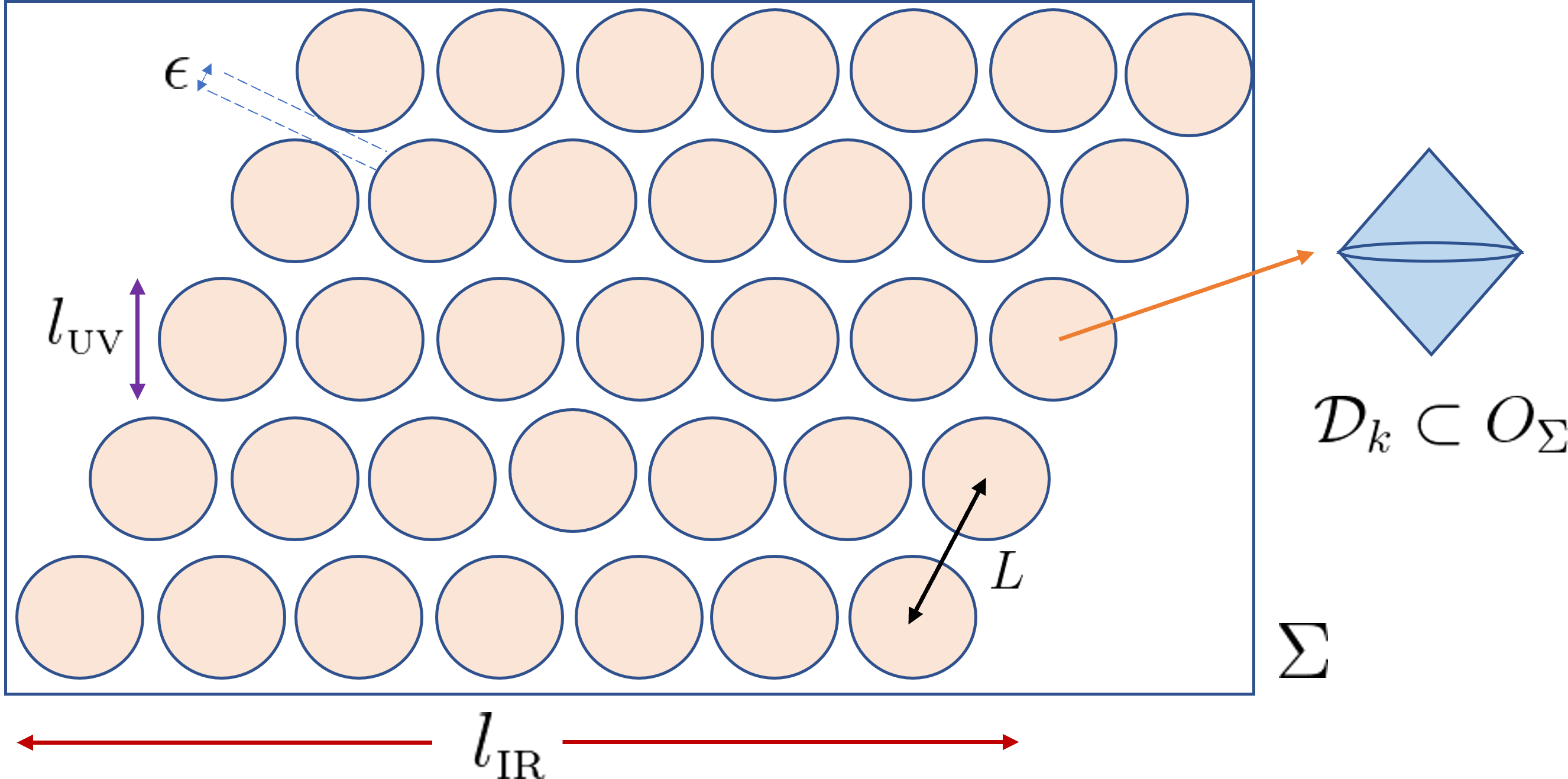}
    \caption{Tiling of the Cauchy surface $\Sigma$ using the supports of the local mode. Given $N$ local modes $(\hat{x}_k,\hat{p}_k)$ such that the supports $\mathcal{D}_k\equiv \supp f^{(k)}_a$ ($a=1,2$) are contained in the thickening of the Cauchy slice $O_\Sigma$. For finite $N$ and compact $\mathcal{D}_k$, one can tile a subset of the Cauchy surface. How large the region is to be covered defines an IR scale, and how small the tiles are defines a UV scale. Another UV scale $\epsilon$ defines a `finite corridor' between each tile.}
    \label{fig: tiling}
\end{figure}

An important feature of this construction is that the size of the supports $\mathcal{D}_k$ gives us an ultraviolet (UV) cutoff, and the number of modes $N$ gives us an infrared (IR) cutoff. This clearly has a natural analogue in the case of harmonic lattice approximation of the scalar field theory (via discretization), but with one important difference: the UV and IR cutoffs in this construction are \textit{covariant} in the sense that it is purely geometrical and does not depend on any coordinate system. Furthermore, instead of discretizing the field theory and losing the symmetries of the continuum limit directly, what we have done is to choose a set of smearing functions with supports that provide a \textit{discrete tiling} $\Sigma$ ({See Figure}~\ref{fig: tiling}). There is an additional UV scale $\epsilon$ that arises from allowing for ``finite corridor'' between each tile, something we allow for in situations where this may be relevant (e.g., for computation of entropic quantities in QFT \cite{hollands2017entanglement}). Since the tiling is geometrical in nature, our construction naturally generalizes to curved spacetimes. It is also worth noting that the appearance of these UV/IR cutoffs in the construction is quite natural and physically motivated. More importantly,  we will show that some of the relativistic symmetries will be preserved in the Wigner function. 

Let $\omega:\A(\M)\to \C$ be an algebraic state  acting on the algebra of observables (see Appendix~\ref{appendix: AQFT}). Essentially, for any observable $A\in \A(\M)$ the state $\omega(A)$ gives the expectation value of the observable $A$. The \textit{Wigner function} $W_\omega$ for a single mode associated with some state $\omega$ is defined to be the expectation value of the displaced parity operator  $\Delta^{(0)}_{H_3}(\alpha)$ \cite{Royer1977wigner,glauber1969kernel,brif1999phase}, given by
\begin{equation}
    \begin{aligned}
    {W}_\omega(\alpha) 
    &\coloneqq \omega(\Delta_{H_3}^{(0)}(\alpha))\equiv \tr\bigr(\hat{\rho}_\omega {\Delta}_{H_3}^{(0)}(\alpha) \bigr) \\
    \Delta^{(0)}_{H_3}(\alpha)&= \hat{D}(\alpha)e^{\ii\pi\hat{a}^\dagger\hat{a}/2}\hat{D}(-\alpha)\,,
    \end{aligned}
\end{equation}
where $\alpha\in \C$, $\hat{\rho}_\omega$ the density operator representation of $\omega$, and 
\begin{align}
    \hat{D}(\alpha) = e^{\alpha\hat{a}^\dagger - \alpha^*\hat{a}}
\end{align}
is the displacement operator. In fact, the operator $\Delta^{(0)}_{H_3}(\alpha)$ is a special case of a more general one-parameter family of operators $\Delta^{(s)}_{G}(\alpha)$, where $s\in [-1,1]$ and $G$ is some dynamical symmetry group, known as the \textit{Stratonovich-Weyl} (SW) \textit{kernel} \cite{brif1999phase}. For the \textit{optical phase space} where the symmetry group is the Heisenberg-Weyl group $H_3$ giving rise to the CCR \eqref{eq: oscillator-CCR}, the SW kernel defines a one-parameter family of quasiprobability distribution with integral representation  given by \cite{glauber1969kernel,brif1999phase}
\begin{align}
    \Delta_{H_3}^{(s)} (\alpha) &= \int_{\C}\frac{\dd^2\xi}{\pi}e^{s|\xi|^2/2}e^{\xi^*\alpha-\xi\alpha^*}{\hat{D}(\xi)}
    \label{eq: stratonovich-weyl-kernel}
\end{align}
Here we have $\dd^2\xi \equiv \dd(\Re\xi)\,\dd(\Im \xi) = \frac{1}{2}\dd\xi\,\dd \xi^*$. We will focus on the case when $s=0$ that corresponds to the Wigner function\footnote{The case for $s=\pm 1$ correspond to the well-known Glauber-Sudarshan $P$ function and the Husimi $Q$ function respectively studied by Cahill and Glauber \cite{glauber1969kernel}. Brif and Mann extended this to more general dynamical symmetry groups in \cite{brif1999phase}. }. 

Observe that the exponent of the displacement operator $\hat{D}(\xi)$ can be re-written as follows. Using the shorthand $R_\xi\equiv \Re\,\xi$ and $I_\xi\equiv \Im\,\xi$, we can write
\begin{align}
    \hat{D}(\xi) = e^{\xi\hat{a}^\dagger-\xi^*\hat{a}} 
    &= e^{\ii\hat{\phi}(h)}\,,
\end{align}
where the complex smearing function $h$ is given by
\begin{align}
    h &\coloneqq \sqrt{\frac{2}{\hbar}}\rr{I_\xi f_1 - R_\xi f_2}\,.
\end{align}
Hence, the SW kernel can be formally written as
\begin{align}
    W_\omega(\alpha) \equiv \int_{\C}\frac{\dd^2\xi}{\pi} e^{\xi^*\alpha-\xi\alpha^*}\omega(e^{\ii \hat{\phi}(h)})\,.
\end{align}
So far we have not done anything conceptually different from the standard phase space quantum mechanics for QHO apart from the fact that the displacement operator now depends on \textit{spacetime smearing function} $h$. Note also that although $\hat{\phi}(h)$ is unbounded, $e^{\ii\hat{\phi}(h)}$ is bounded. 

The simplest example is given in the realm of ``Gaussian quantum information'' \cite{adesso2014gaussian,weedbrook2012gaussian} where we consider \textit{Gaussian states}. For instance, if the field is in a \textit{quasifree state}, i.e., Gaussian with vanishing one-point functions, the Wigner function is given by (see Appendix~\ref{appendix: AQFT})
\begin{align}
    W_\omega(\alpha) = \int_{\C}\frac{\dd^2\xi}{\pi}e^{\xi^*\alpha-\xi\alpha^*}e^{-\mathsf{W}(h,h)/2}\,,
\end{align}
where
\begin{align}
    \mathsf{W}(h,h) &= \frac{2}{\hbar} R_\xi^2\mathsf{W}(f_2,f_2) + \frac{2}{\hbar} I_\xi^2\mathsf{W}(f_1,f_1) \notag\\
    &\hspace{2cm} + \frac{4}{\hbar} I_\xi R_\xi \Re[\mathsf{W}(f_1,f_2)]\,.
\end{align}
Hence the Wigner function is Gaussian since $\mathsf{W}(h,h)$ is a quadratic function of $\xi$.  This can be seen clearer in the `position-momentum' coordinates: in the $(q,p)$ coordinates, we have \cite{glauber1969kernel}
\begin{align}
    \frac{\dd^2\xi}{\pi}&\to\frac{\dd q'\dd p'}{2\pi\hbar }\,,\quad 
    \xi \to \frac{1}{\sqrt{2\hbar}}( q'+\ii p')\,.
\end{align}
This gives
\begin{align}
    W_\omega(q,p) &= \int_{\R^2}\frac{\dd q'\dd p'}{2\pi\hbar }
    e^{\frac{\ii(pq'-qp')}{2\pi}}e^{-\mathsf{W}(h,h)/2}\,,
\end{align}
but now we have\footnote{This also makes clear that $\mathsf{W}(h,h)$, hence the Wigner function, is \textit{invariant} under simple rescaling $(f_1,f_2)\to (\lambda f_1,\lambda^{-1}f_2)$.}
\begin{align}
    \mathsf{W}(h,h) &= \frac{( q')^2}{2\pi^2}\mathsf{W}(f_2,f_2)+\frac{(p')^2}{2\pi^2}\mathsf{W}(f_1,f_1)\notag\\
    &\hspace{2cm}+\frac{q'p'}{\pi^2}\Re\,{\mathsf{W}(f_1,f_2)}
    \label{eq: wightman-h}
\end{align}
which is clearly quadratic in $q'$ and $p'$. Note that the smeared two-point functions $\mathsf{W}(f_a,f_b)$ are all \textit{fixed} constants once we fix the definition of the local mode, so $\mathsf{W}(h,h)$ is really a polynomial in $(\xi,\xi^*)$ or $(q',p')$.

In order to generalize the Wigner function to include $N$ local modes, it is useful to adopt the abstract index notation and the Einstein summation convention typically used in relativity (although not common in QIT). We follow the notation 
 in \cite{EMM2022wigner} with some small modifications. For a single local mode, we first define the ``Darboux coordinates'' $\xi^a\equiv (x,p)$, $\eta^a\equiv (x',p')$ and hence we can rewrite the Wigner function as
\begin{align}
    W_\omega(x,p) &= \frac{1}{2\pi\hbar}\int \dd^2\eta \,e^{\ii\eta^a\Omega_{ab}\xi^b}\omega\bigr(e^{\ii \eta^a\Omega_{ab}\hat{\xi}^b}\bigr)
\end{align}
where $\hat{\xi}^b\equiv (\hat{x},\hat{p})$ and $\Omega\equiv \Omega_{ab}$ is the \textit{symplectic two-form} of the phase space. Since by construction the local mode is defined in the Darboux coordinates, the matrix elements of $\Omega$ is given by $\Omega_{ab} = E(f_a,f_b)$, i.e., 
\begin{align}
    \Omega\equiv \begin{bmatrix}
        0 & 1 \\ -1 & 0 
    \end{bmatrix}\,.
\end{align}
Writing $\Omega(\eta,\xi)\coloneqq \Omega_{ab}\eta^a\xi^b$ and using the shorthand $\mathsf{f}\coloneqq (f_1,f_2)$, we can now write
\begin{align}
    \Omega(\eta,\hat\xi) &= x'\hat{p}-p'\hat{x} \equiv \hat{\phi}(\mathsf{h}_{\eta})\,,
\end{align}
where $\mathsf{h}_{\eta}\coloneqq \Omega(\eta,\mathsf{f}) = x'f_2-p'f_1$. Consequently, we have
\begin{align}
    W_\omega(\xi) &= \frac{1}{2\pi\hbar}\int\dd^2\eta \,e^{\ii\Omega(\eta,\xi)}\omega(e^{\ii\hat{\phi}(\mathsf{h}_{\eta})})\,.
    \label{eq: Wigner-symplectic}
\end{align}
In the literature, the integral of the form \eqref{eq: Wigner-symplectic} is also called the \textit{symplectic Fourier transform} with respect to the characteristic function (see, e.g., \cite{gross2006wigner}) 
\begin{align}
    \chi(\eta) &\coloneqq \omega(e^{\ii\hat{\phi}(\mathsf{h}_\eta)})\equiv \braket{e^{\ii\hat{\phi}(\mathsf{h}_\eta)}}_\omega\,.
\end{align}
However, in the context of QFT it has an additional interpretation: it says that $\chi(\eta)$ is the expectation value of some displacement operator  \textit{localized in spacetime}!

The generalization of the Wigner function to $N$ local modes now follows straightforwardly. We extend $\xi^a,\eta^a$ to $2N$ variables, i.e., 
\begin{equation}
    \begin{aligned}
    \xi^a \equiv (x_1,p_1,x_2,p_2,...,x_N,p_N)\,,\\
    \eta^a \equiv (x_1',p_1',x_2',p_2',...,x_N',p_N')\,,
    \end{aligned}
\end{equation}
and correspondingly the operator version $\hat{\xi}^a$. The joint symplectic two-form is now given by
\begin{align}
    {\Omega} = \bigoplus_{k=1}^N{\Omega}^{(k)}\,,\quad {\Omega}^{(k)} = \begin{bmatrix}
        0 & 1 \\ -1 & 0
    \end{bmatrix}\,,
\end{align}
where the matrix elements of each block are obtained from the fact that we have ${\Omega}^{(k)}_{ab} = E(f_{a\phantom{b}}^{(k)},f_b^{(k)})$. We can now write the multimode Wigner function as
\begin{align}
    W^{(N)}_\omega(\xi) &= {\frac{1}{(2\pi\hbar)^N}}\int\dd^{2N}\!\eta \,e^{\ii\Omega(\eta,\xi)}\chi(\eta)\,,
    \label{eq: Wigner-symplectic_multimode}
\end{align}
where $\chi(\eta)=\omega\bigr(e^{\ii\hat{\phi}(\mathsf{h}_{\eta})}\bigr)$ is now the multimode characteristic function with the smearing function $\mathsf{h}_\eta$ extended to $2N$ variables:
\begin{align}
    \mathsf{h}_{\eta} &= \Omega(\eta,\mathsf{f})\,,\quad \mathsf{f}\coloneqq (f^{(1)}_1,f^{(1)}_{2},...,f^{(N)}_1,f^{(N)}_{2})\,.
\end{align}
Note that Eq.~\eqref{eq: Wigner-symplectic_multimode} is identical to \eqref{eq: Wigner-symplectic} except that it involves higher-dimensional integrals, as we expect from a ``covariant'' definition.

It is straightforward to check that for quasifree states, the Wigner function is still Gaussian with the expression in Eq.~\eqref{eq: wightman-h} generalized to a homogeneous polynomial of $\eta^a$ with degree 2. Furthermore, by going to the Hilbert space representation of $\omega$ using the Gelfand-Naimark-Segal (GNS) representation theorem (see, e.g., \cite{fewster2019algebraic}), we can in principle calculate the Wigner function for \textit{any} state of interest. For example, to calculate the Wigner function for one-particle Fock state (in the vacuum representation), we have
\begin{align}
    \ket{1_G}\coloneqq\int\dd^n\bk\,G(\bk)\hat{a}_\bk^\dagger\ket{0}\,,
\end{align}
where $G\in L^2(\R^3)$ is square-integrable and $\ket{0}$ is the vacuum state (the GNS vector of the the algebraic vacuum state $\omega_0$). 
Essentially, what is computable in the standard CV settings for finitely many oscillators will in principle be computable in our setting, since the only QFT component is the calculation of the smeared Wightman functions. More generally, beyond Gaussian settings we will need to know the smeared $2N$-point functions for the state $\omega$, given by (see Appendix~\ref{appendix: AQFT})
\begin{equation}
    \begin{aligned}    &\mathsf{W}\bigr(f_1^{(1)},f_2^{(1)},...,f^{(2N)}_1,f^{(2N)}_2\bigr)\notag\\
    &\coloneqq\omega\bigr(\hat{\phi}(f_1^{(1)})\hat\phi(f_2^{(1)})...\hat{\phi}(f_1^{(2N)})\hat{\phi}(f_2^{(2N)})\bigr)\,.
    \end{aligned}
    \label{eq: 2N-point-functions}
\end{equation}
For the case when $\omega$ corresponds to the Poincar\'e-invariant vacuum in Minkowski spacetime, notice that the entire Wigner function is \textit{invariant} under the action of the Poincar\'e group simply because the vacuum Wightman $2N$-point functions \eqref{eq: 2N-point-functions} will be Poincar\'e-invariant.

\section{Discussion and outlook}
In this work we have given a construction of the Wigner function for a bosonic scalar field theory that comes with natural ultraviolet UV and IR scales. Our construction is valid in any $(n+1)$-dimensional globally hyperbolic curved spacetime $\M$ with metric $g_{ab}$. We do not employ the path integral formalism, and neither do we obtain the Wigner function by discretizing the quantum field theory into a lattice bosonic theory. We achieved this by building $N$ local modes from $2N$ smeared field operators, such that they tile a Cauchy surface $\Sigma$ of the spacetime where the field theory lives. The construction can be viewed as a form of ``covariant discretization'' of the quantum field, since the tiling of the Cauchy surface does not depend on any choice of coordinate systems or spacetime foliation. The difficulty of computing the Wigner function for fixed $N$ then reduces to the computation of finitely many smeared correlation functions of the quantum field. This also implies that the Wigner function can inherit the symmetries of the QFT states which are not preserved by lattice discretization.

One of the main takeaway of this work is that the Wigner function is essentially that of a finite number of local modes with fixed UV/IR scales, hence all standard tools from continuous-variable QIT should apply essentially verbatim. Thus we are naturally led to several questions, some of which (not exhaustive) we give below:
\begin{enumerate}[leftmargin=*,label=\arabic*),nolistsep]
    \item How does the continuum limit ($N\to\infty$,  $l_{\textsc{uv}},\epsilon\to 0$) compare with the one obtained from path integrals and lattice discretization? What happens to the Wigner function in the thermodynamic limit $N\to \infty$ (but with fixed UV scales)? Note that this latter case amounts to having countably many QHOs with countably infinite-dimensional optical phase space.

    \item We have not answered the question of how the Wigner function would depend on how we ``tile'' the Cauchy surface. This freedom already exists in the lattice discretization, where the lattice choice is not unique. Are the freedom in choosing the lattice synonymous to how we tile the Cauchy slice?

    \item For fixed tiling parameter list $\mathcal{S}\coloneqq (N,l_{\textsc{uv}},l_{\textsc{ir}},\epsilon)$, we can now compute \textit{Wigner negativity} for the quantum field. How does this compare with the standard CV computation involving real QHOs instead of local modes? How does Wigner negativity vary with $\mathcal{S}$? Does this provide a measure of non-classicality and contextuality for QFT?

    \item How does the Wigner function generalize to other QFTs, e.g., spin-1/2 (fermion), spin-1 (photon), and spin-2 (graviton) fields?

    \item Can the Wigner function be made to work with bulk-to-boundary correspondence in algebraic QFT (see, e.g., \cite{Dappiaggi2005rigorous-holo,Moretti2005BMS-invar,tjoa2022holography}? In the context of AdS/CFT correspondence, could Wigner functions also admit some holographic duality?

    \item What about the Wigner function interacting QFTs such as quantum electrodynamics (QED) in curved spacetime? How does the perturbative quantum  corrections modify the Wigner function?

    \item Perhaps most importantly, how does the Wigner function evolve in time? Since the construction involves UV/IR cutoff, it is not clear that the time evolution of the Wigner function constructed this way is consistent with the field equation of the scalar QFT. 
    
    \item More speculatively, can we construct a `virtual' harmonic lattices, or even tensor networks, using local modes? Can we assign meaning to the `virtual' local Hamiltonian constructed out of the local modes? 
\end{enumerate}
It is worth emphasizing that these questions apply also to the other one-parameter family of quasiprobability distributions defined through the SW kernel \eqref{eq: stratonovich-weyl-kernel}. 

The margin of this paper is too small to address all the above questions, so we will leave them for further work.

\section*{Acknowledgment}

E.T. thanks Valerio Scarani for the hospitality at the Centre for Quantum Technologies (CQT), Singapore, where this work could be completed. E.T. also thanks Bruno de S. L. Torres, Robie A. Hennigar, Robert B. Mann, Koji Yamaguchi, and Jack Davis for useful discussions. This work was supported in part by the Natural Sciences and Engineering Research Council of
Canada (NSERC).

\bibliography{wigner-ref}

\appendix

\section{Scalar QFT in curved spacetimes}
\label{appendix: AQFT}

In this section we briefly review the algebraic framework for quantization of a real scalar field in an arbitrary (globally hyperbolic) curved spacetimes. We will follow closely the summary in \cite{tjoa2022holography},  which in turn is based on a very well-written exposition by \cite{KayWald1991theorems}. An accessible introduction to $*$-algebras and $C^*$-algebras for the algebraic formulation of quantum mechanics (AQM) and quantum field theory (AQFT) can be found in \cite{fewster2019algebraic}. We have also benefitted from the discussions about the local modes from the very clear writing in \cite{Ruep2021harvesting}, which also covers the quantization of scalar field  complementary to the exposition in \cite{fewster2019algebraic,Khavkhine2015AQFT,KayWald1991theorems}.

\subsection{Algebra of observables and algebraic states}

Let $\phi$ be a free, real scalar field in an $(n+1)$-dimensional globally hyperbolic spacetime $(\mathcal{M},g_{ab})$ with metric tensor $g_{ab}$. The scalar field is said to be a \textit{Klein-Gordon field} if it obeys the Klein-Gordon equation
\begin{align}
     P\phi = 0\,,\quad  P = \nabla_a\nabla^a - m^2  - \xi R\,,
     \label{eq: KGE}
\end{align}
where $\xi \geq 0$, $R$ is the Ricci scalar and  $\nabla$ is the Levi-Civita connection with respect to $g_{ab}$. Global hyperbolicity ensures that $\M$ admits a foliation by spacelike Cauchy surfaces $\Sigma_t$ labelled by real parameter $t\in \R$, thus Eq.~\eqref{eq: KGE} is well-posed as a Cauchy problem. 

Let $f\in \CS$ be a smooth compactly supported test function on $\M$. Let $E^\pm(\sx,\sy)$ be the retarded and advanced propagators associated with the Klein-Gordon operator $P$, such that
\begin{align}
    E^\pm f\equiv (E^\pm f)(\sx) \coloneqq \int \dd V'\, E^\pm (\sx,\sx')f(\sx') \,,
\end{align}
solves the inhomogeneous equation $P(E^\pm f) = f$. Here $\dd V' = \dd^n\sx'\sqrt{-g}$ is the invariant volume element. The \textit{causal propagator} is defined to be the advanced-minus-retarded propagator $E=E^--E^+$. It is known that if $O$ is an open neighbourhood of some Cauchy surface $\Sigma$ and $\varphi$ is any real solution with compact Cauchy data to Eq.~\eqref{eq: KGE}, denoted $\varphi \in \Sol_\R(\M)$, then there exists $f\in \CS$ with $\supp(f)\subset O$ such that $\varphi=Ef$ \cite{Khavkhine2015AQFT}.

The quantization of $\phi$ can be regarded as an $\R$-linear mapping from the space of smooth compactly supported test functions to a unital $*$-algebra $\A(\M)$, i.e., 
\begin{align}
    \hat\phi: C^\infty_0(\mathcal{M})&\to \A(\M)\,,\quad f\mapsto \hat\phi(f)\,,
\end{align}
which obeys the following conditions:
\begin{enumerate}[leftmargin=*,label=(\alph*)]
    \item (\textit{Hermiticity}) $\hat\phi(f)^\dag = \hat\phi(f)$ for all $f\in \CS$;
    \item (\textit{Klein-Gordon}) $\hat\phi(Pf) = 0$ for all $f\in \CS$;
    \item (\textit{Canonical commutation relations}  ) We have $[\hat\phi(f),\hat\phi(g)] = \ii E(f,g)\openone $ for all $f,g\in \CS$, where $E(f,g)$ is the smeared causal propagator
    \begin{align}
        E(f,g)\coloneqq \int \dd V f(\sx) (Eg)(\sx)\,.
    \end{align}
    \item (\textit{Time slice axiom}) Let $\Sigma\subset \M$ be a Cauchy surface and $O$ a fixed open neighbourhood of $\Sigma$. $\A(\M)$ is generated by the unit element $\openone$ (hence $\A(\M)$ is unital) and the smeared field operators $\hat\phi(f)$ for all $f\in \CS$ with $\supp(f)\subset O$.
\end{enumerate}
The $*$-algebra $\A(\M)$ is called the \textit{algebra of observables} of the real Klein-Gordon field. The \textit{smeared} field operator reads
\begin{align}\label{eq: ordinary smearing}
    \hat\phi(f) = \int \dd V\hat\phi(\sx)f(\sx)
    \,,
\end{align}
with $\hat\phi(\sx)$ seen as an operator-valued distribution. 

Next, given  $\Sol_\R(\M)$ --- the vector space of real-valued solutions of \eqref{eq: KGE} with compact Cauchy data --- we can promote it to a \textit{symplectic} vector space by equipping it with a symplectic form $\sigma:\Sol_\R(\M)\times\Sol_\R(\M)\to \R$, defined as
\begin{align}
    \sigma(\phi_1,\phi_2) \coloneqq \int_{\Sigma_t}\!\! {\dd\Sigma^a}\,\Bigr[\phi_{{1}}\nabla_a\phi_{{2}} - \phi_{{2}}\nabla_a\phi_{{1}}\Bigr]\,,
    \label{eq: symplectic form}
\end{align}
where $\dd \Sigma^a = -t^a \dd\Sigma$, $-t^a$ is the inward-directed unit normal to the Cauchy surface $\Sigma_t$, and $\dd\Sigma = \sqrt{h}\,\dd^3\bx$ is the induced volume form on $\Sigma_t$ \cite{Poisson:2009pwt,wald2010general}. This is independent of the Cauchy surface, and we can use this to view $\hat\phi(f)$ as a \textit{symplectically smeared field operator}  \cite{wald1994quantum} 
\begin{align}
    \label{eq: symplectic smearing}
    {\hat\phi(f) \equiv \sigma(Ef,\hat\phi)\,,}
\end{align}
and the CCR algebra can be written as 
\begin{align}
    {[\sigma(Ef,\hat\phi),\sigma(Eg,\hat\phi)] = \ii\sigma(Ef,Eg)\openone = \ii E(f,g)\openone \,,}
\end{align}
where $\sigma(Ef,Eg) = E(f,g)$ in the second equality follows from Eq.~\eqref{eq: ordinary smearing} and \eqref{eq: symplectic smearing}. 

It is often more convenient (though not quite necessary for our results in this work) to work with the ``exponentiated'' version of $\A(\M)$ called the \textit{Weyl algebra} --- denoted $\W(\M)$ --- since its elements are (formally) bounded operators. The Weyl algebra $\W(\M)$ is a unital $C^*$-algebra generated by the elements which formally take the form 
\begin{align}
    W(Ef) \equiv 
    {e^{\ii\hat\phi(f)}}\,,\quad f\in \CS\,.
    \label{eq: Weyl-generator}
\end{align}
These elements satisfy \textit{Weyl relations}:
\begin{equation}
    \begin{aligned}
    W(Ef)^\dagger &= W(-Ef)\,,\\
    W(E (Pf) ) &= \openone\,,\\
    W(Ef)W(Eg) &= e^{-\frac{\ii}{2}E(f,g)} W(E(f+g))
    \end{aligned}
    \label{eq: Weyl-relations}
\end{equation}
where $f,g\in \CS$. The characteristic function of the Wigner function, as we saw earlier, is essentially given as an expectation value of some Weyl generator with a fixed smearing function $\mathsf{h}_\eta$.


Given $\A(\M)$, a quantum state for the theory is called an \textit{algebraic state}: it is defined as  a $\C$-linear functional $\omega:\A(\M)\to \C$ such that 
\begin{align}
    \omega(\openone) = 1\,,\quad  \omega(A^\dagger A)\geq 0\quad \forall A\in \A(\M)\,.
    \label{eq: algebraic-state}
\end{align}
In effect $\omega$ takes as input an observable and outputs expectation values: Eq.~\eqref{eq: algebraic-state} says that the state is normalized to unity and positive-semidefinite operators have non-negative expectation values. The state $\omega$ is pure if it cannot be written as $\omega= \alpha \omega_1 + (1-\alpha)\omega_2$ for any $\alpha\in (0,1)$ and any two distinct algebraic states $\omega_1,\omega_2$, otherwise it is mixed.

The standard Hilbert space representation (i.e., the Fock representation) in canonical quantization can be obtained from the Gelfand-Naimark-Segal (GNS) reconstruction theorem \cite{wald1994quantum,Khavkhine2015AQFT,fewster2019algebraic}: that is, we can construct a \textit{GNS triple}\footnote{We may also need a dense subset $\mathcal{D}_\omega\subset \mathcal{H}_\omega$ for $\A(\M)$.} $(\mathcal{H}_\omega, \pi_\omega,{\ket{0_\omega}})$, where $\pi_\omega: \mathcal{\A(\M)}\to {\text{End}(\mathcal{H}_\omega)}$ is a Hilbert space representation with respect to state $\omega$ such that any algebraic state $\omega$ can be realized as a \textit{vector state} {$\ket{0_\omega}\in\mathcal{H}_\omega$}. The observables $A\in \A(\M)$ are then represented as operators $\hat A\coloneqq \pi_\omega(A)$ acting on the Hilbert space. With the GNS representation, the action of algebraic states is given by $\omega(A) = \braket{0_\omega|\hat A|0_\omega}$. Since in the context of QFT in curved spacetimes we have  infinitely many unitarily inequivalent representations of the CCR algebra, the algebraic framework provides a unifying procedure to quantize the scalar field theory.

\subsection{Quasifree states}

In AQFT, all physically reasonable algebraic states are believed to be within the subclass known as \textit{Hadamard states} \cite{Khavkhine2015AQFT,KayWald1991theorems,Radzikowski1996microlocal}: roughly speaking, they are states with the correct singularity structure. A state is said to be \textit{quasifree} if they are completely characterized by their two-point correlation functions (in CV terminology, these are Gaussian states centered at the origin in phase space). The vacuum states and (squeezed) thermal states are quasifree states, while coherent states are Gaussian (but not quasifree) states. Some authors do not distinguish quasifree states and simply regard all of them as a subclass of Gaussian states. For us, what matters is that if a state is quasifree, then the expectation value of the Weyl generator is Gaussian (see \cite{tjoa2022holography,Ruep2021harvesting,KayWald1991theorems,Khavkhine2015AQFT,fewster2019algebraic} for details):
\begin{align}
    \omega (W(Ef))\equiv \omega(e^{\ii\hat{\phi}(f)}) = e^{-\mathsf{W}(f,f)/2}\,,
    \label{eq: quasifree}
\end{align}
where $\mathsf{W}(f,f)$ is the symmetrically-smeared Wightman two-point functions associated to the quasifree state. 

We can calculate $\mathsf{W}(f,f)$ most straightforwardly using the canonical quantization procedure. There, the real scalar field $\phi$ is promoted to an operator-valued distribution $\hat\phi$ which can be written in terms of the Fourier mode decomposition
\begin{align}
    \hat\phi(\sx) &= \int \dd^n\bk \rr{a_\bk u_\bk(\sx)+a_\bk^\dagger u_\bk^*(\sx)}\,,
\end{align}
where $\{u_\bk(\sx)\}$ are positive-frequency modes (with respect to some timelike vector fields) of Klein-Gordon operator $P$. The Klein-Gordon inner product is given by complex extension of the symplectic form on the \textit{complexified} space of solutions $\Sol_\C(\M)$:
\begin{align}
    (\phi_1,\phi_2)_{\textsc{kg}} &\coloneqq \ii\sigma (\phi^*_1,\phi_2) \,.
    \label{eq: KG-inner-product}
\end{align}
where $\dd \Sigma^a = -t^a \dd\Sigma$, $-t^a$ is the inward-directed unit normal to the Cauchy surface $\Sigma_t$, and $\dd\Sigma = \sqrt{h}\,\dd^3\bx$ is the induced volume form on $\Sigma_t$ \cite{Poisson:2009pwt,wald2010general}. The mode functions are normalized with respect to the Klein-Gordon inner product:
\begin{equation}
    \begin{aligned}
    (u_\bk,u_{\bk'})_\textsc{kg}  &= \delta^n(\bk-\bk')\,,\quad (u_\bk^{\phantom{*}},u^*_{\bk'})_\textsc{kg} = 0\,,\\
    (u_\bk^*,u^*_{\bk'})_\textsc{kg} &= -\delta^n(\bk-\bk')\,.
    \end{aligned}
    \label{eq: KG-normalization}
\end{equation}
For example, standard calculations (see, e.g., \cite{birrell1984QFTCS}) show that the unsmeared vacuum Wightman two-point function can be computed in terms of the mode functions:
\begin{align}
    \mathsf{W}(\sx,\sy) &= \int \dd^n\bk\, u^{\phantom{*}}_\bk(\sx) u^*_\bk(\sy)\,,
\end{align}
from which we can calculate the symmetrically smeared two-point function as
\begin{align}
    \mathsf{W}(f,f) = \int \dd V\,\dd V' f(\sx)f(\sy)\mathsf{W}(\sx,\sy)\,.
    \label{eq: Wightman-double-smeared}
\end{align}
Generalizations to other non-vacuum states can be made, e.g., by using the annihilation or creation operators acting on the Fock vacuum.

\end{document}